\documentclass{elsart3-1}

\usepackage[english,francais]{babel}

\newtheorem{e-proposition}[theorem]{Proposition}

\newtheorem{e-definition}[theorem]{Definition\rm}

\renewcommand{\theequation}{\arabic{equation}}
\def\og{\leavevmode\raise.3ex\hbox{$\scriptscriptstyle\langle\!\langle$~}}
\def\fg{\leavevmode\raise.3ex\hbox{~$\!\scriptscriptstyle\,\rangle\!\rangle$}}

\def \D {\hbox{d}}

\def \PVI    {{\rm P6}}
\def\xI{x}
\def\tI{t}
\def\XI{X}
\def\TI{T}
\def\xR{\tI}
\def\tR{\xI}
\def\XR{\TI}
\def\TR{\XI}
\def\KRiccati{\vartheta} 
\def\CRAS{C.~R.~Acad.~Sc.~Paris}
\def\AnnENS{Ann.~\'Ec.~Norm.~}

\journal{the Acad\'emie des sciences}
\begin{document}
\centerline{}
\begin{frontmatter}


  \title{The master Painlev\'e VI heat equation}


\selectlanguage{english}
\author[al1,al2,al3]{Robert Conte},
\ead{Robert.Conte@cea.fr}
\author[al2]{Ivan Dornic}
\ead{Ivan.Dornic@cea.fr}

\address[al1]{LRC MESO, 
    Centre de math\'ematiques et de leurs applications (UMR 8536) 
\\ et CEA-DAM, \'Ecole normale sup\'erieure de Cachan, 
\\ 61, avenue du Pr\'esident Wilson, F--94235 Cachan Cedex, France.}
\address[al2]{Service de physique de l'\'etat condens\'e (CNRS URA 2464)
\\Orme des merisiers, CEA-Saclay, F--91191 Gif-sur-Yvette Cedex, France.}
\address[al3]{Department of Mathematics, The University of Hong Kong,
\\ Pokfulam Road, Hong Kong.}


\medskip
\begin{center}
{\small Received *****; accepted after revision +++++\\
Presented by £££££}
\end{center}

\begin{abstract}
\selectlanguage{english}
Given the second order scalar Lax pair of the sixth Painlev\'e equation,
we build a generalized heat equation with rational coefficients
which does not depend any more on the Painlev\'e variable.
{\it To quote 
 this article: 
R.~Conte, I.~Dornic, C.~R.~Acad.~Sci.~Paris, Ser.~I ??? (201x).}

\vskip 0.5\baselineskip

\selectlanguage{francais}
\noindent{\bf R\'esum\'e} \vskip 0.5\baselineskip \noindent
{\bf L'\'equation ma\^\i tresse de la chaleur associ\'ee \`a Painlev\'e VI}

\'Etant donn\'e la paire de Lax scalaire de la sixi\`eme \'equation de 
Painlev\'e,
nous donnons une construction directe de l'\'equation de
la chaleur g\'en\'eralis\'ee \`a coefficients rationnels
qui ne d\'epend plus de la variable de Painlev\'e.
{\it Pour citer cet article~: 
R.~Conte, I.~Dornic, C.~R.~Acad.~Sci.~Paris, Ser.~I ??? (201x).}
\end{abstract}
\end{frontmatter}


\selectlanguage{francais}
\section*{Version fran\c{c}aise abr\'eg\'ee}

Soit l'\'equation diff\'erentielle ordinaire (EDO) (\ref{eqFuchsPbODE}),
dot\'ee de quatre singularit\'es fuchsiennes $\xI=\xI_{\nu}=\infty,0,1,\tI$
et d'une singularit\'e apparente $\xI=u$.
La condition d'isomonodromie 
(ind\'ependance de la monodromie envers le birapport $\tI$
des quatre singularit\'es fuchsiennes)
\'equivaut \`a une condition diff\'erentielle entre $u$ et $\xR$,
qui a ainsi conduit R.~Fuchs \cite{FuchsP6}
\`a la d\'ecouverte de la sixi\`eme \'equation de Painlev\'e \PVI\ 
(\ref{eqPVI}).

Le processus d'isomonodromie conduit \`a adjoindre \`a l'EDO lin\'eaire 
(\ref{eqFuchsPbODE})
une deuxi\`eme \'equation lin\'eaire, 
ce couple (\ref{eqGarnierP6LaxODE})--(\ref{eqGarnierP6LaxPDE})
d\'efinissant en langage moderne une paire de Lax scalaire.

Le but de cet article est de donner 
pour la premi\`ere fois
une preuve constructive de l'existence d'une \'equation de la chaleur
g\'en\'eralis\'ee, voir (\ref{eqheatg}),
dont les coefficients sont ind\'ependants de la fonction de Painlev\'e $u$
et ne d\'ependent, sous une forme rationnelle,
que du birapport $\tI$ et de la variable $\xI$.
Les deux d\'emonstrations ant\'erieures 
n'\'etaient valides, comme d\'etaill\'e section \ref{SectionFinale},
que sous la condition 
$(\theta_\infty^2,\theta_{0}^2,\theta_{1}^2,\theta_{\xR}^2)
 \not=(1,1,1,1)$,
voir relation (\ref{asths}).

\selectlanguage{english}

\section{Introduction. The scalar Lax pair of the sixth Painlev\'e equation}

Let us first recall the 1905 classical result of R.~Fuchs \cite{FuchsP6}.
Consider a second order linear ordinary differential equation (ODE) 
for a function $\psi=\psi(\xI)$ 
--- the wave function, or wave vector ---
with four Fuchsian singularities of crossratio $\tI$,
put for convenience (but without loss of generality after a homographic transformation) 
at $\xI=\xI_{\nu}=\infty,0,1,\tI$. 
As prescribed by Poincar\'e to have sufficient degrees of freedom 
for the isomonodromy problem to be non-trivial \cite[pp.~217--220]{Poincare1883},
 one must in addition put 
 one apparent singularity located at $\xI=u$, 
so that the ODE satisfied by $\psi$ writes \cite[Eq.~(1)]{FuchsP6}:
\begin{eqnarray}
\frac{\D^2 \psi}{\D \xI^2}
& - & 
\left[
  \frac{A}{ \xI   ^2}
+ \frac{B}{(\xI-1)^2}
+ \frac{C}{(\xI-\tI)^2}
+ \frac{E}{(\xI-u)^2} 
+ \frac{a}{ \xI}
+ \frac{b}{ \xI-1}
+ \frac{c}{ \xI-\tI   }
+ \frac{e}{ \xI-u   }
\right] \psi=0.
\label{eqFuchsPbODE}
\end{eqnarray}
In Eq.~(\ref{eqFuchsPbODE}),
$A,B,C,E$ are constant parameters (independent of $\tI$ and $\xI$), while 
$a,b,c,e$ will ultimately depend on $\tI$ but not on $\xI$.
The requirement that the monodromy matrix
(which transforms two independent solutions $\psi_1,\psi_2$
when $\xI$ goes around a singularity $\xI_{\nu}$)
be independent of the location of the nonapparent singularity $\tI$
--- the isomonodromy condition --- 
results in the constraint that $u$, as a function of the deformation parameter $\tI$,
obeys  the nonlinear second order ordinary differential equation,
\begin{eqnarray}
\frac{\D^2 u}{\D \tI^2}
&=&
\frac{1}{2} \left[\frac{1}{u} + \frac{1}{u-1} + \frac{1}{u-\tI} \right] 
\left(\frac{\D u}{\D \tI}\right)^2
- \left[\frac{1}{\tI} + \frac{1}{\tI-1} + \frac{1}{u-\tI} \right] 
\frac{\D u}{\D \tI}
\nonumber \\ & &
+ \frac{u (u-1) (u-\tI)}{\tI^2 (\tI-1)^2}
  \left[\alpha + \beta \frac{\tI}{u^2} + \gamma \frac{\tI-1}{(u-1)^2}
        + \delta \frac{\tI (\tI-1)}{(u-\tI)^2} \right].
        \label{eqPVI}
\end{eqnarray}
Eq.~(\ref{eqPVI}) is the celebrated sixth Painlev\'e equation \PVI, 
the most general second order nonlinear ODE without movable critical singularities,
which lies at the crossroads of many problems of mathematics and theoretical
physics of current active interest.
The set of four parameters $(\alpha, \beta, \gamma, \delta)$ 
is in one-to-one correspondence with 
the set $A,B,C,E$ defined in (\ref{eqFuchsPbODE})
and with the squares $\theta_{\nu}^2$ of the monodromy exponents,
see relations (\ref{asths}) and (\ref{eqasAs}) below.

Coming back to Eq.~(\ref{eqFuchsPbODE}),
what happens more precisely is that demanding the isomonodromy of Eq.~(\ref{eqFuchsPbODE})
 is tantamount to the existence
of two linear equations for the wave vector (now a function $\psi=\psi(\xI,\tI)$).
To endorse a modern terminology, the corresponding Fuchs-Garnier {\it scalar Lax pair} of equations writes 
 \cite{FuchsP6,Fuchs1907,GarnierThese}: 
\begin{eqnarray}
& &
\partial^2_{\xI} \psi + (S/2) \psi=0,\
\label{eqGarnierP6LaxODE} 
\\
& &
\partial_{\tI} \psi + W \partial_{\xI} \psi -(1/2) W_{\xI} \psi = 0,\
\label{eqGarnierP6LaxPDE} 
\end{eqnarray}
and their commutativity (or compatibility) condition 
yields \PVI. 
In their most concise form (first exhibited by Garnier), 
the two scalar functions $S,W$ 
display a remarkable symmetry between $\tR$ and $u$:
\begin{eqnarray}
& &
{\hskip -5.0 truemm}
-\frac{S}{2}=
\frac{3/4}{(\tR-u)^2}
+ \frac{g_1 u' + g_0}{(\tR-u) \tR(\tR-1)} 
+\frac{[(g_1 u')^2 - g_0^2] \displaystyle\frac{u-\xR}{u (u-1)}
+ f_{\rm G}(u)}{\tR(\tR-1)(\tR-\xR)}
+ f_{\rm G}(\tR),
\label{eqS}
\\ & & {\hskip -5.0 truemm}
W=-\displaystyle\frac{\tR(\tR-1) (u-\xR)}{(\tR-u) \xR(\xR-1)},
  \quad
g_1=\displaystyle-\frac{\xR (\xR-1)}{2 (u-\xR)},
  \quad
g_0=-u+\frac{1}{2},\
\\ & & {\hskip -5.0 truemm}
f_G(z)=\displaystyle\frac{A}{z^2} + \frac{B}{(z-1)^2}
\displaystyle+\frac{C}{(z-\xR)^2} + \frac{E}{z (z-1)},
\\ & & {\hskip -5.0 truemm}
\label{eqasAs}
(2\alpha, -2\beta, 2\gamma, 1-2\delta) =(4(A+B+C+E+1),4 A+1,4 B+1,4 C+1).
\end{eqnarray}

The purpose of this paper is to eliminate the dependent variable $u$ (and its derivative)
between the two linear equations 
(\ref{eqGarnierP6LaxODE})--(\ref{eqGarnierP6LaxPDE})
while preserving the linearity of the resulting single equation. 
 This provides us with a heat equation for the wave vector 
whose coefficients are solely rational functions of $\tI$ and $\xI$ (and of the monodromy parameters $\theta_{\nu}$).
This generalized heat equation 
had in fact appeared earlier in the literature. 
We shall discuss this in the final section of the present Note.
The paper is organized as follows.
In section \ref{SectionElimination},
we present the elimination procedure.
In section \ref{SectionFinale},
we compare our findings with previous results, discussing in particular 
the underlying motivation.

\section{From the scalar Lax pair to the generalized heat equation}
\label{SectionElimination}

The guideline of our procedure 
is the singularity structure of the two linear equations 
(\ref{eqGarnierP6LaxODE})--(\ref{eqGarnierP6LaxPDE})
in the complex plane of $\xI$.
To achieve our goal, 
it is necessary (but not sufficient) to eliminate the polar singularity $\xI =u$
between the two equations.

The first equation (\ref{eqGarnierP6LaxODE}) is an ODE with five Fuchsian singularities 
in the complex plane of $\xI$ whose Riemann scheme is
\begin{eqnarray}
& & {\hskip -8.0truemm}
\pmatrix{
\infty & 0 & 1 & \xR & u \cr 
 (1-\theta_\infty)/2 & (1-\theta_0)/2 & (1-\theta_1)/2 & (1-\theta_\xR)/2 & -1/2 \cr 
 (1+\theta_\infty)/2 & (1+\theta_0)/2 & (1+\theta_1)/2 & (1+\theta_\xR)/2 &  3/2 \cr 
},
\end{eqnarray}
with the correspondence
\begin{eqnarray}
& & 
(2 \alpha,-2 \beta,2 \gamma,1-2 \delta)
=(\theta_\infty^2,\theta_{0}^2,\theta_{1}^2,\theta_{\xR}^2).
\label{asths}
\end{eqnarray}
As to the second equation (\ref{eqGarnierP6LaxPDE}),
its ODE reduction $\partial_{\tR}=0$ possesses one Fuchsian singularity at $\tR=u$,
with the Riemann scheme
\begin{eqnarray}
& & {\hskip -8.0truemm}
\pmatrix{
u \cr 
 -1/2 \cr 
}.
\end{eqnarray}

In a first step, we remove the four finite double poles in (\ref{eqGarnierP6LaxODE}),
\textit{via} the change of wave function
\begin{eqnarray}
& & 
\psi=\tR^{(1-\theta_0)/2} (\tR-1)^{(1-\theta_1)/2} (\tR-\tI)^{(1-\theta_\xR)/2}
     (\tR-u)^{-1/2} e^{G(\xR)} \Psi,
\label{eqpsi-to-Psi0}
\end{eqnarray}
in which the gauge 
 $G$ is a function of $\tI$ which is left for the moment arbitrary. 
The change of wave function (\ref{eqpsi-to-Psi0}) is quite similar to the
classical one for the Gauss hypergeometric equation.
After decomposition of its coefficients in simple elements of $\tR$,
the Lax pair becomes
\begin{eqnarray}
& & {\hskip -5.0truemm}
\partial_\tR^2 \Psi 
+ \left(\frac{1-\theta_0}{\tR}+\frac{1-\theta_1}{\tR-1}
       +\frac{1-\theta_\xR}{\tR-\xR} -\frac{1}{\tR-u} \right) 
  \partial_\tR \Psi 
\nonumber\\ & & {\hskip -5.0truemm} \phantom{\partial_\tR^2 \Psi }
+ \frac{1}{4 u(u-1)(u-\xR)} \left(\frac{R_0}{\tR}+\frac{R_1}{\tR-1}
 +\frac{R_\xR}{\tR-\xR}+\frac{2 R_u}{\tR-u} \right) \Psi=0,\
\label{eqlax2} 
\\
& & {\hskip -5.0truemm}
\xR (\xR-1) \partial_\xR \Psi 
 -\frac{\tR(\tR-1)(u-\xR)}{\tR-u} \partial_\tR \Psi 
  + \left(\xR (\xR-1) G'+\frac{R_u}{2 (\tR-u)} 
             +\frac{(\theta_0+\theta_1+\theta_\xR-1)(u-\xR)}{2}
\right)\Psi=0,\
\label{eqlax1} 
\end{eqnarray}
in which the residues $R_j$ 
are best expressed in terms of 
the two relations defining the one-parameter classical Riccati solution
of \PVI\ in terms of the hypergeometric function \cite{Fuchs1907},
\begin{eqnarray}
& & 
R(\theta_0,\theta_1,\theta_\xR) \equiv
\xR(\xR-1)u'+u(u-1)(u-\xR)
   \left(\frac{\theta_0}{u}+\frac{\theta_1}{u-1}+\frac{\theta_\xR-1}{u-\xR} \right)=0,\
\\ & &
\KRiccati  \equiv (1-\theta_0-\theta_1-\theta_\xR)^2-\theta_\infty^2=0.
\end{eqnarray}
These residues are
\begin{eqnarray}
& & 
R_u=R(\theta_0,\theta_1,\theta_\xR),
\nonumber\\ & & 
R_0  =-\frac{R(\theta_0,\theta_1,\theta_\xR) R(2-\theta_0, -\theta_1, -\theta_\xR)
    +\KRiccati u(u-1)(u-\xR) u}{\xR},
\nonumber\\ & & 
R_1  =-\frac{R(\theta_0,\theta_1,\theta_\xR) R( -\theta_0,2-\theta_1, -\theta_\xR)
    +\KRiccati u(u-1)(u-\xR) (u-1)}{(1-\xR)},
\nonumber\\ & & 
R_\xR=-\frac{R(\theta_0,\theta_1,\theta_\xR) R( -\theta_0, -\theta_1,2-\theta_\xR)
    +\KRiccati u(u-1)(u-\xR) (u-\xR)}{\xR(\xR-1)}.
\nonumber\\ & & 
\end{eqnarray}
It is remarkable that $-R_\xR/(4 u(u-1)(u-\xR))$ 
is precisely the polynomial Hamiltonian of \PVI\ \cite{MalmquistP6}.

In a second step, we eliminate this simple pole. 
Since the quotient of the residues 
of (\ref{eqlax2}) and (\ref{eqlax1}) at the simple pole $\tR=u$ 
does not depend on $\Psi$, 
the resulting equation remains linear in $\Psi$, 
\begin{eqnarray}
& & {\hskip 0truemm}
-\frac{\xR(\xR-1)}{\tR(\tR-1)(\tR-\xR)}\partial_{\xR} \Psi 
+
\partial_{\tR}^2 \Psi 
-\left(\frac{\theta_0-1}{\tR}+\frac{\theta_1-1}{\tR-1}+\frac{\theta_\xR}{\tR-\xR} \right)
 \partial_\tR \Psi 
\nonumber\\ & & {\hskip 0truemm} \phantom{-\xR(\xR-1)\partial_\xR \Psi}
+ \frac{1}{\tR(\tR-1)(\tR-\xR)}
  \left[\frac{\KRiccati}{4}(\tR-\xR)-\xR(\xR-1) G'(\xR)-F(\xR) \right] \Psi=0,
\label{eqheatG} 
\end{eqnarray}
and its dependence on $u$ and $u'$ is gathered 
in an expression independent of $t$,
\begin{eqnarray}
\label{toto}
& & {\hskip 0truemm}
F(\xR)=-\frac{R(\theta_0,\theta_1,\theta_\xR)R(-\theta_0,-\theta_1,-\theta_\xR)} 
             {4 u(u-1)(u-\xR)}
+\left(\theta_\infty^2+1-(\theta_0+\theta_1+\theta_\xR)^2\right)
\frac{u-\xR}{4}.
\end{eqnarray}
The third and last step is to choose the arbitrary function $G(\xR)$
so as to cancel this
contribution of $u$ and $u'$. 
The final result is a generalized heat equation 
whose coefficients are rational functions of $\xR$ and $\tR$,
\begin{eqnarray}
& & {\hskip -11.0truemm}
-\xR(\xR-1)\partial_\xR \Psi 
+\tR(\tR-1)(\tR-\xR) \left[
\partial_\tR^2 \Psi 
-\left(\frac{\theta_0-1}{\tR} 
      +\frac{\theta_1-1}{\tR-1}
			+\frac{\theta_\xR}{\tR-\xR} \right)
 \partial_\tR \Psi 
\right]
+ \left[\frac{\KRiccati}{4}(\tR-\xR)-g(\xR) \right] \Psi=0,\
\label{eqheatg} 
\end{eqnarray}
in which $g(\xR)$ can be arbitrarily chosen. 
In the Picard case $\theta_\nu=0$, its reduction $\partial_\xR=0, g(\xR)=0$ 
is identical to the classical linear ODE of Legendre 
for the periods of the elliptic function.


\section{Discussion}
\label{SectionFinale}

The heat equation we have obtained is in fact not new,
but the present proof is the first one without any restriction.
Indeed,
previous occurrences of this heat equation are the following.
\begin{enumerate}

\item
By establishing a formal correspondence between the scalar Lax pair
(\ref{eqGarnierP6LaxODE})--(\ref{eqGarnierP6LaxPDE})
and the time-dependent Schr\"odinger equation of quantum mechanics,
Suleimanov \cite{Suleimanov1994} obtained this heat equation.

\item
Starting from the second order matrix Lax pair of \PVI\
as given by Jimbo and Miwa \cite{JimboMiwaII},
in which the monodromy matrix is just the sum of four simple poles,
D.P.~Novikov \cite{N2009} assumed, like Ref.~\cite{JimboMiwaII},
that the residue at $\infty$ is a constant matrix
and finally proved that the first component of the two-dimensional
wave vector obeys the heat equation (\ref{eqheatg}).

\item
In a context of quantization of classical integrable systems 
(like Ref.~\cite{Suleimanov1994}),
Zabrodin and Zotov \cite{ZZ2012}
also started from the matrix Lax pair of Ref.~\cite{JimboMiwaII}
and, 
after a suitable gauge transformation and change of variables,
obtained a master \PVI\ heat equation of the form 
\begin{equation}
\partial_{\XR} \Psi = (1/2)\partial^2_{\TR} \Psi+V(\TR,\XR) \Psi. 
\end{equation}
This time-dependent Schr\"odinger (or Fokker-Planck) equation coincides 
with the rational heat equation (\ref{eqheatg})
after a point transformation 
$\xR \to \XR=\XR(\xR), \tR \to \TR=\TR(\tR,\xR)$ 
involving elliptic functions
totally analogous to the transformation \cite{FuchsP6,PaiCRAS1906}
which maps \PVI\ to a Hamiltonian
which is the sum of a kinetic energy and a time-dependent potential energy.

\end{enumerate}

The drawback with the derivations of Refs \cite{N2009} and \cite{ZZ2012}
is that, because of the assumption made in Ref.~\cite{JimboMiwaII}
that the residue at $\infty$ is a constant matrix,
the matrix Lax pair as assumed by Jimbo and Miwa does not exist
when all four $\theta_\nu^2$ are unity,
see details in \cite{LCM2003,C2006Kyoto}.

The earliest occurrence of the heat equation (\ref{eqheatg})
which we are aware of 
is in conformal field theory \cite[Eq.~(5.17)]{BPZ}
in the particular case of a value $c=1$ of the central charge
of a Virasoro algebra.

{}From the present results one deduces easily by the classical confluence
of the four singularities \cite{PaiCRAS1906,CMBook}
similar results for all the other Painlev\'e functions.
In particular,
the Tracy-Widom probability distribution in random matrix theory
\cite{TW1994a}
has been recently characterized by such a 
time-dependent Schr\"odinger equation
\cite{BV2013},
associated to the second Painlev\'e function.
Similar results hold for the sine-Gordon third Painlev\'e function 
\cite{DornicP3SG}.





\vfill\eject
\begin{thebibliography}{99}

\bibitem{BPZ} A.A.~Belavin, A.M.~Polyakov and A.B.~Zamolodchikov, 
Infinite conformal symmetry in two-dimensional quantum field theory,
Nucl.~Phys.~B {\bf 241} (1984) 333--380.

\bibitem{BV2013} A.~Bloemendal and B.~Vir\'ag, 
Limits of spiked random matrices $I$, 
Proba.~theory rel.~fields {\bf 156} (2013) 795--825.
http://arxiv.org/abs/1011.1877

\bibitem{C2006Kyoto} R.~Conte,
On the Lax pairs of the sixth Painlev\'e equation,
RIMS K\^oky\^uroku Bessatsu {\bf B2} (2007) 21--27.
http://arXiv.org/abs/nlin.SI/0701049

\bibitem{CMBook} R.~Conte and M.~Musette,
{\it The Painlev\'e handbook} (Springer, Berlin, 2008).
Russian translation
{\it Metod Penleve y ego prilozhenia} 
(Regular and chaotic dynamics, Moscow, 2011).

\bibitem{DornicP3SG} I.~Dornic,                                         
Phase-noise distribution, Brownian motion in time-dependent potentials, and the Sine-Gordon 
Painlev\'e III transcendent,
in preparation. 

\bibitem{FuchsP6} R.~Fuchs,                                         
Sur quelques \'equations diff\'erentielles lin\'eaires du second ordre,
\CRAS\ {\bf 141} (1905) 555--558. 

\bibitem{Fuchs1907} R.~Fuchs,                                    
\"Uber lineare homogene Differentialgleichungen zweiter Ordnung mit drei im
Endlichen gelegenen wesentlich singul\"aren Stellen,
Math.~Annalen {\bf 63} (1907) 301--321.

\bibitem{GarnierThese} R.~Garnier,                         
Sur des \'equations diff\'erentielles du troisi\`eme ordre dont
l'int\'e\-gra\-le
g\'en\'erale est uniforme et sur une classe d'\'equations nouvelles d'ordre
sup\'e\-ri\-eur dont l'int\'egrale g\'en\'erale a ses points critiques fixes,
\AnnENS {\bf 29} (1912) 1--126.

\bibitem{JimboMiwaII} M.~Jimbo and T.~Miwa,
Monodromy preserving deformations of linear ordinary differential equations
with rational coefficients.~II,
Physica D {\bf 2} (1981) 407--448.

\bibitem{LCM2003} R.~Lin, R.~Conte and M.~Musette,
On the Lax pairs of the continuous and discrete sixth Painlev\'e equations,
J.~Nonlinear Mathematical Physics {\bf 10}, Supp.~2, 107--118 (2003).

\bibitem{MalmquistP6} J.~Malmquist,
Sur les \'equations diff\'erentielles du second ordre dont l'int\'egrale
g\'en\'erale a ses points critiques fixes,
Arkiv f\"or Math.~Astr.~Fys.~{\bf 17} (1922--23) 1--89.

\bibitem{N2009} D.P.~Novikov,                                   
The 2x2 matrix Schlesinger system and the Belavin--Polyakov--Za\-mo\-lod\-chi\-kov
 system,
Teoreti\-cheskaya i Matematicheskaya Fizika {\bf 161} (2009) 191--203.
                         Theor.~Math.~Phys.~{\bf 161} (2009) 1485--1496.

\bibitem{PaiCRAS1906} P.~Painlev\'e,
Sur les \'equations diff\'erentielles du second ordre \`a points critiques
fixes,
\CRAS\ {\bf 143} (1906) 1111--1117. 

\bibitem{Poincare1883} H.~Poincar\'e,                        
Sur les groupes des \'equations lin\'eaires,
Acta mathematica {\bf 4} (1883) 201--312.
Reprinted, {\it O$\!$euvres} (Gauthier-Villars, Paris, 1951--1956),
tome II, 300--401.

\bibitem{Suleimanov1994} B.I.~Suleimanov,                           
Hamiltonian property of the Painlev\'e equations and the method of 
isomonodromic deformations,
Differentsial'nye Uravneniya {\bf 30} (1994) 791--796
       [English~: Diff.~equ.~{\bf 30} (1994) 726--732].

\bibitem{TW1994a} C.A.~Tracy and H.~Widom,                          
Level-spacing distributions and the Airy kernel,
Commun.~Math.~Phys.~{\bf 151} (1994) 151--174.

\bibitem{ZZ2012} A.~Zabrodin and A.~Zotov,
Quantum Painlev\'e-Calogero correspondence,
J.~Math.~Phys.~{\bf 53}, 073507 (2012);
Quantum Painlev\'e-Calogero correspondence for Painlev\'e $\rm{VI}$,
J.~Math.~Phys.~{\bf 53}, 073508 (2012),
http://arxiv.org/abs/1107.5672

\end{thebibliography}
\end{document}